\documentstyle[aps,preprint]{revtex}

\begin{document}

\draft

\title{Heavy quark supermultiplet excitations}

\author{R. Delbourgo and Dongsheng Liu\cite{Add}}

\address{Physics Department, University of Tasmania,
         Hobart, Australia 7005}

\date{9 December 1993}

\maketitle

\begin{abstract}
Lorentz-covariant wave functions for meson and baryon supermultiplets are
simply derived by boosting SU(2)$_{\rm spin}$ representations corresponding
to multiquark systems at rest.
\end{abstract}

\section{Introduction}

A number of articles have recently appeared \cite{IWetal} where descriptions
of hadronic excitations containing one or more heavy quarks have been
presented.  All of these recent constructions have regarded the external
velocity of the hadron $v$ as a four-vector parameter and have written
the wave functions in terms of $v$ and of internal velocities $u$; in these
schemes one encounters projections parallel and perpendicular to $v$ and
tensors/spinors formed out of $v$, the several $u$ and the gamma matrices.
Some of the descriptions are quite elaborate and involve complicated
algebraic manoeuvres over Lorentz covariant objects before the final forms
are attained. Occasionally the derivations are given in the Bethe-Salpeter
framework \cite{BS} or variants thereof.

Most of these authors were (understandably) unaware that this problem was
tackled many years ago \cite{DS} in the context of Reggeization of
supermultiplet theory. At that time there was interest in constructing
supermultiplet wavefunctions at arbitrary integer total angular momentum $J$
(or its Casimir generalization for the chosen supersymmetry group),
before continuing somehow to complex $J$-values. Multispinor wavefunctions
appearing in section IV of reference 3 were quoted abruptly without much
elaboration and the formulae may therefore look rather mysterious today.
Because this subject of hadronic excitations has come back into vogue
in the context of heavy quark physics, we shall explain here the derivation
of multiquark excitation functions and will take the opportunity to correct
a few normalization factors in reference 3. Then we shall go on to discuss
the application to heavy quark composites.

The basic idea is extremely simple and direct: since all such
supermultiplets are massive we can always proceed to the rest frame,
where the four-velocity $v$ is a unit timelike vector pointing along
the time axis. In that frame the only surviving space-time symmetry is
the little group of the full Lorentz group associated with $v$, namely
spatial SU(2), and the states fall into irreducible representations of it.
We now know (as was not fully appreciated in the pre-$bc$ year 1969) that
hadrons are {\em white} composites of quarks and gluons; the standard picture
views the mesons as made of one quark and an antiquark, while baryons are
made of three quarks, plus a colourless mixture of any number of gluons and
quark-antiquark pairs. Some models sometimes replace two quasi-free quarks
by a single composite diquark, but the basic idea is essentially the same.
The internal degrees of freedom and the binding mechanism---which is not
completely understood at low energies---produce a bound multiquark state
that carries appropriate quantum numbers; these can be nothing more
than the SU(2)$_J$ labels plus a possible multiplicity label $N$ for states
which are repeated, as well as the internal flavour group quantum numbers
which we have not bothered to expose. In other words the integrations
over the internal momenta eventually lead, in the rest frame, to meson
and baryon states
\begin{equation}
 [\phi_a^{\bar{b}}]_{\{m_1..m_L\}}^N \quad {\rm and} \quad
   [\psi_{abc}]_{\{m_1..m_L\}}^N,
\end{equation}
where $m$ is a O(3) vector index corresponding to orbital excitation $L$,
and $a,b,c$ stand for two-component spinor indices (barred for antiquarks).
\{..\} represents a symmetrised tensor product and we have assumed above
that all Kronecker traces over the $m$ indices are zero to make the orbital
state irreducible with respect to O(3). In the next section we shall reduce
these states with respect to total angular momentum $J$ and in the
following section we will boost up the results to arbitrary velocity $v$.
Finally we shall discuss the connection with Lagrangians, Bethe-Salpeter
wavefunctions and other work.

\section{Non-relativistic reduction into $J$ representations}

The first task is to simplify the spin structure. Because the quark and
antiquarks can in principle be acted upon by different spin groups
(see section IV), we first reduce the multispinors into total spin states,
disregarding their orbital quantum numbers,
$$ \phi_a^{\bar{b}} = [\phi_5\delta_a^{\bar{b}} +
                       (\sigma_m)_a^{\bar{b}}\phi_m]/\sqrt{2}, $$
$$\psi_{abc} = \psi_{\{abc\}}+\psi_{\{ab\}c}+\psi_{[ab]c}
             = (\sigma_2\sigma_m)_{\{ab\}} \psi_{mc}/\sqrt{2} +
               (\epsilon_{ac}\psi_b + \epsilon_{bc}\psi_a)/\sqrt{6}
               + \epsilon_{ab}\psi_c'/\sqrt{2}. $$
Since antiparticles have opposite parity to particles we recognise the
pseudoscalar state $\phi_5 (0^{--})$, the vector state $\vec{\phi}$
($1^{--}$), the spin 1/2 states $\psi,\psi'$ and the spin 3/2 state
$\vec{\psi}_c$, obeying the irreducibility condition,
$\sigma_m\psi_m = 0$ which ensures that $\psi_{\{abc\}}$ is symmetric.
Note the occurrence of the $\sigma_2$ matrix which is the lowering operator
for SU(2) and the charge conjugation matrix non-relativistically.

The next step is to combine spin and orbital factors into representations
of total spin $J$. As far as the pseudoscalar meson excitations are
concerned, there is nothing more to be done: $\phi_{5\{m_1..m_L\}}$ stands
for a state with parity $(-1)^{L+1}$ and $CP=1$; the vector meson
excitations $(1\times L)$ require reduction to states with $J = L+1,L,L-1$
as given below:
\begin{eqnarray}
\phi_{m\{m_1..m_L\}}&=&\phi^{(L+1)}_{\{mm_1\cdots m_L\}}+\frac{1}{\sqrt{2}L}
    \sum_k i\epsilon_{mm_kn}\phi^{(L)}_{\{m_1\cdots\bar{k}\cdots m_Ln\}}
     \nonumber \\
       &+&\frac{1}{L}\sqrt{\frac{2L-1}{2L+1}} \sum_k \left[
       \delta_{mm_k} \phi^{(L-1)}_{\{m_1\cdots\bar{k}\cdots m_L\}} -
       \frac{2}{2L-1} \sum_l \delta_{m_km_l}\phi^{(L-1)}_{\{mm_1\cdots
       \bar{k}\bar{l}\cdots m_L\}} \right];
\end{eqnarray}
a bar over an O(3) index, like $\bar{k}$, signifies that $m_k$ is missing
from the tensor.

Turning to the baryons, one either needs to combine the orbital momentum
with spin 1/2 or with spin 3/2. In the former case one arrives at states
with $J=L+1/2$ and $J=L-1/2$, encapsulated by the decomposition
\begin{equation}
 \psi_{\{m_1..m_L\}} = \psi^{(L+1/2)}_{\{m_1\cdots m_L\}} +
      \frac{1}{\sqrt{L(2L+1)}}\sum_k \sigma_{m_k}
      \psi^{(L-1/2)}_{\{m_1\cdots\bar{k}\cdots m_L\}},
\end{equation}
while in the latter case one must reduce to $J=L+3/2,L+1/2,L-1/2$ and $L-3/2$
representations:
$$ \psi_{m\{m_1..m_L\}}  = \psi^{(L+3/2)}_{\{mm_1\cdots m_L\}} + $$
$$ \sqrt{\frac{3}{4L(2L+3)}}\left[ \sum_k (
    i\epsilon_{mm_kn}\psi^{(L+1/2)}_{\{m_1\cdots\bar{k}\cdots m_Ln\}}
    +\sigma_{m_k}\psi^{(L+1/2)}_{\{mm_1\cdots\bar{k}\cdots m_L\}}) -
    \frac{L}{3}\sigma_m\psi^{(L+1/2)}_{\{m_1\cdots m_L\}} \right]  + $$
$$ \sqrt{\frac{3(2L-1)}{L(L+1)(2L+1)}} \sum_k \left[ (
    \frac{2\delta_{mm_k}}{3} - \frac{i\epsilon_{mm_kn}\sigma_n}{3})
    \psi^{(L-1/2)}_{\{m_1\cdots\bar{k}\cdots m_L\}} - \sum_l
    \frac{2\delta_{m_km_l}}{2L-1}
    \psi^{(L-1/2)}_{\{mm_1\cdots\bar{k}\bar{l}\cdots m_L\}} \right] + $$
\begin{equation}
 \frac{1}{L\sqrt{(L-1)(2L+1)}} \sum_{k,l} \left[
    (\delta_{mm_k}\sigma_{m_l} - \frac{2\sigma_m\delta_{m_km_l}}{2L-1})
    \psi^{(L-3/2)}_{\{m_1\cdots\bar{k}\bar{l}\cdots m_L\}} -
    \sum_n \frac{2\delta_{m_km_l}\sigma_{m_n}}{2L-1}
    \psi^{(L-3/2)}_{\{mm_1\cdots\bar{k}\bar{l}\bar{n}\cdots m_L\}} \right].
\end{equation}
These expressions are correctly normalized, in as much as  $1 = \sum_J
|\psi^{(J)}_{\{m_1\cdots\}}|^2 $, like the lefthand sides of Eqs. (2) to (4).

\section{Boosted wavefunctions}

It is not widely appreciated that Lorentz covariant expressions for particle
wavefunctions are readily obtained by boosting the non-relativistic formulae.
An incoming meson which is a composite of an incoming quark and antiquark
must contain the projection factors $[(1+\gamma_0)/2]\Gamma[(1-\gamma_0)/2]$
in the rest frame in order to pick out the upper two components of the quark
and the lower two components of the other quark. Likewise the
non-relativistic expressions $(\sigma_2)_{[ab]}$ and
$(\sigma_2\sigma_m)_{\{ab\}}$ should be interpreted as the upper
$2\times 2$ components of the four-component multispinors
$[(1+\gamma_0)\gamma_5 C/2]_{\alpha\beta}$ and
$[(1+\gamma_0)\gamma_m C/2]_{\alpha\beta}$.
{}From this point of view it is easy to understand why the properly
boosted versions (any direction of $v$) of the wavefunctions (2) are
\begin{equation}
 [\phi(v)]_\alpha^\beta=[(1 + \gamma.v)(\gamma_5 \phi_5(v)
         -\gamma^{\mu}\phi_{\mu}(v))/2\sqrt{2}]_\alpha^\beta,
\end{equation}
where vector indices on fields are orthogonal to $v$, or
$v^\mu\phi_{\mu}= 0$, signifying that the $\phi$ are polarization vectors.
Also the generalization of the non-relativistic condition on the
vector-spinor, $\vec{\sigma}.\vec{\psi}=0$ reads $\gamma.\psi=v.\psi=0$.
Results of this type were originally derived \cite{DSS} by carrying out
Lorentz-covariant reductions of multispinors, using Bargmann-Wigner
equations acting on each Dirac spinor index and solving the constraint and
symmetry conditions. (They have been rediscovered several times in different
ways.) In the light of experience this is an unnecessarily
complicated way of proceeding: one simply ``solves'' the equations
\cite{DRSS} in the rest frames, a totally trivial step, and boosts up
to arbitrary $v$, as above. The only other substitutions we must be careful
with are for spin matrices and the Kronecker delta,
$$\sigma_m \rightarrow v^\nu\sigma_{\nu\mu}\gamma_5 \equiv w_\mu, \qquad
  \delta_{mn}\rightarrow -\eta_{\mu\nu}+ v_\mu v_\nu \equiv d_{\mu\nu}(v),$$
where we have assumed that $v$ has unit length on-shell.

With this point made, we may readily understand why the relativistic
versions of Eqs. (2), (3) and (4) are
\begin{eqnarray}
\phi_{\mu\{\mu_1..\mu_L\}}&=&\phi^{(L+1)}_{\{\mu\mu_1\cdots \mu_L\}}
    + \frac{1}{\sqrt{2}L} \sum_k
    iv^{\lambda}\epsilon_{\lambda\mu\mu_k\nu}d^{\nu\nu'}
    \phi^{(L)}_{\{\mu_1\cdots\bar{k}\cdots\mu_L\nu'\}} \nonumber \\
       &+&\frac{1}{L}\sqrt{\frac{2L-1}{2L+1}} \sum_k \left[
       d_{\mu\mu_k} \phi^{(L-1)}_{\{\mu_1\cdots\bar{k}\cdots \mu_L\}} -
       \frac{2}{2L-1} \sum_l d_{\mu_k\mu_l}\phi^{(L-1)}_{\{\mu\mu_1\cdots
       \bar{k}\bar{l}\cdots\mu_L\}} \right];
\end{eqnarray}
\begin{equation}
 \psi_{\{\mu_1..\mu_L\}} = \psi^{(L+1/2)}_{\{\mu_1\cdots\mu_L\}} +
      \frac{1}{\sqrt{L(2L+1)}}\sum_k w_{\mu_k}
      \psi^{(L-1/2)}_{\{\mu_1\cdots\bar{k}\cdots \mu_L\}};
\end{equation}
$$ \psi_{\mu\{\mu_1..\mu_L\}} = \psi^{(L+3/2)}_{\{\mu\mu_1\cdots\mu_L\}} +$$
$$ \sqrt{\frac{3}{4L(2L+3)}}\left[ \sum_k(iv^{\lambda}
    \epsilon_{\lambda\mu\mu_k\nu} d^{\nu\nu'}
    \psi^{(L+1/2)}_{\{\mu_1\cdots\bar{k}\cdots\mu_L\nu'\}}
    +w_{\mu_k}\psi^{(L+1/2)}_{\{\mu\mu_1\cdots\bar{k}\cdots\mu_L\}}) -
    \frac{L}{3}w_\mu\psi^{(L+1/2)}_{\{\mu_1\cdots\mu_L\}} \right]  +$$
$$ \sqrt{\frac{3(2L-1)}{L(L+1)(2L+1)}} \sum_k \left[ (
    \frac{2d_{\mu\mu_k}}{3}-\frac{iv^{\lambda}
    \epsilon_{\lambda\mu\mu_k\nu}w^{\nu}}{3})
    \psi^{(L-1/2)}_{\{\mu_1\cdots\bar{k}\cdots\mu_L\}} - \sum_l
    \frac{2d_{\mu_k\mu_l}}{2L-1}
    \psi^{(L-1/2)}_{\{\mu\mu_1\cdots\bar{k}\bar{l}\cdots\mu_L\}} \right] +$$
\begin{equation}
 \frac{1}{L\sqrt{(L-1)(2L+1)}} \sum_{k,l} \left[
    (d_{\mu\mu_k}w_{\mu_l} - \frac{2w_\mu d_{\mu_k\mu_l}}{2L-1})
    \psi^{(L-3/2)}_{\{\mu_1\cdots\bar{k}\bar{l}\cdots\mu_L\}} -
    \sum_n \frac{2d_{\mu_k\mu_l}w_{\mu_n}}{2L-1}
    \psi^{(L-3/2)}_{\{\mu\mu_1\cdots\bar{k}\bar{l}\bar{n}\cdots\mu_L\}}
    \right].
\end{equation}

These expressions make no reference to the internal momenta, nor should
they. The final covariant wavefunctions carry the quantum numbers
associated with external momentum and its little group, no more and no less.
For instance, the first excited states ($L=1$) of mesons ($0^{++}, 1^{-+},
2^{++}$) and baryons ($1/2 ^-,3/2^-, 5/2^-$) are represented by
\begin{equation}
 \phi_{\mu\nu} = \phi_{\{\mu\nu\}}^{(2)} -\frac{i}{\sqrt{8}}v^\lambda
   \epsilon_{\lambda\mu\nu\kappa}\eta^{\kappa\kappa'}\phi_{\kappa'}^{(1)}
   + \frac{1}{\sqrt{3}}d_{\mu\nu}\phi^{(0)};\quad
 \psi_\mu = \psi_\mu^{(3/2)} + w_\mu\psi^{(1/2)}/\sqrt{3} ;
\end{equation}
$$\psi_{\mu\nu} = \psi_{\{\mu\nu\}}^{(5/2)} + \sqrt{\frac{3}{20}}\left[
  -iv^\lambda\epsilon_{\lambda\mu\nu\kappa}\eta^{\kappa\kappa'}
  \psi_{\kappa'}^{(3/2)} + w_\nu\psi_\mu^{(3/2)} - \frac{1}{3}
  w_\mu\psi_\nu^{(3/2)} \right] +\frac{1}{3\sqrt{2}}\left[
  2d_{\mu\nu} - iv^\lambda\epsilon_{\lambda\mu\nu\kappa}w^\kappa\right]
  \psi^{(1/2)}. $$
The only place where a connection may be made with the internal, relative
momenta---and the dynamics that leads to such bound states---is via the
dependence of the masses on total angular momentum, $p^2 = M_J^2(N)$, and on
the (suppressed) quantum number $N$, differentiating between states of the
same $J$. We shall turn to this aspect of the problem now.

\section{Connection with Lagrangians and bound state equations}

It will pay us to re-examine the origin of supermultiplet symmetries as
the older interpretation \cite{DSS} differs slightly from the modern
viewpoint advocated by the Mainz and Harvard schools \cite{MH}. Consider a
set of quark fields with different masses described by the Lagrangian,
\begin{equation}
 {\cal L} = \sum_{F,C} \int d^4x \; \bar{Q}_{FC}(x)[\gamma.(i\partial
            - gA(x)) - m_F]Q_{FC}(x) + {\cal L}_A + {\cal L}_{\rm ew}
\end{equation}
where $F$ stands for the flavour and $C$ for colour and $A$ is the
(matrix-valued) gluon octet field. If we disregard gluon interactions and
make a Fourier transformation to momentum space we get
\begin{equation}
 {\cal L}_{\rm free} = \sum_{F,C} \int d^4p \; \bar{q}_{FC}(p)[\gamma.p
                      - m_F]q_{FC}(p).
\end{equation}
The early relativistic interpretations of the Wigner supermultiplet symmetry
for particle physics\cite{GR} neglected mass differences between quarks
in order to show that the momentum space wavefunctions admitted an
$SU(2N_F) \times SU(2N_F)$ invariance of ${\cal L}$. It was Isgur and Wise
\cite{IW} who realised that this assumption of mass equality was not needed
and who extended the concept to heavy quarks like $b$ and $c$. Heeding
their lesson, let us change to velocity space by redefining
$h_{FC}(v) = M^{5/2}q(p)$ as a new heavy quark field (normalised to
1 compared with $q(p)$ which was normalised to $M^{-5/2}$). The free
part of the Lagrangian
\begin{equation}
 {\cal L} = \sum_{F,C} \int d^4v \; \bar{h}_{FC}(v)[\gamma.v - 1]h_{FC}(v)
           + {\cal L}_{\rm int}(A) + {\cal L}_{\rm ew},
\end{equation}
then admits a $U(2N_F)\times U(2N_F)$ symmetry between the quark fields
at fixed velocity. This is most readily seen in the rest frame
$v_0 = (1,0,0,0) $ with $h(v_0)$ obeying the projection equation
$(\gamma_0 - 1)h(v_0) = 0$. The little group generators then consist of
\begin{equation}
 [1, \gamma_0, \vec{\sigma}, \vec{\sigma}\gamma_0]\quad\times\quad T_F^{F'},
\end{equation}
where $T$ are the generators of the $U(N_F)$ flavour algebra. When boosted
to any velocity $v$,
\begin{equation}
 S(L_v)\gamma_0 S^{-1}(L_v) = \gamma.v ,
\end{equation}
the free spinors $h(v) = S(L_v)h(v_0)$ retain that supersymmetry but we
should now interpret the generators as
\begin{equation}
 [1, v.\gamma, w_\mu, w_\mu v.\gamma] \quad\times\quad T_F^{F'}.
\end{equation}
Note that we have not toed the modern party line which considers the heavy
quark field $h$ to be a full function of space-time $x$ with $v$
regarded as an external parameter (eventually identified with the velocity
of the heavy hadron, one we may need to integrate over at the end of the
day \cite{Georgi}). Instead we have surmised that $v$ is associated with
the Fourier transform space, so that integration over $v$ is automatic
and {\em $x$ does not appear at all}. The similarity transformation $S(L_v)$
which carries spinors from the rest frame $v_0$ to any $v$ can be construed
as the Fourier transform of a coordinate-space Foldy-Wouthuysen-like
transformation \cite{Melosh} applying to $h(x)$.

The usefulness of this concept hinges upon consideration of the quark-gluon
interaction
$${\cal L}_{\rm int} = \sum_{C,F}\int d^4v\,d^4v'\;
                        g\bar{h}(v)A(v-v').Th(v')/m_F. $$
where the fluctuations over the gluon field $A$ produce a distribution over
momenta of order $\Lambda_{\rm QCD}$. Since this will lead to corrections
$\Lambda_{\rm QCD}/m_F$ to the free system, these will be most
substantial for the lighter quarks and will become negligible as the quark
mass $m_F$ gets ever larger, which Isgur and Wise \cite{IW} first observed.
Thus one concludes that the really significant supersymmetry when QCD becomes
operational is for the heavy flavours. However we believe that the
formulation above, where the heavy quarks fields are functions of
four-velocity $v$ alone and not $x$ and $v$ simultaneously, brings this
feature out much more clearly and elegantly. It may be possible to carry out
a further transformation,
$$ Q(x) \rightarrow T[\exp i\int_{-\infty}^x A(\xi).d\xi]Q(x) $$
in order to simplify the quark-gluon interaction (at the expense of the
measure in the functional integral over the fields), but that is irrelevant
for the present discussion.

To finish off let us briefly compare our formulation of hadronic excitations
with the work of others \cite{IWetal},\cite{BS}. The early work by Isgur
{\em et al} concerning strong and semileptonic decays of excited hadrons
was cast in a non-covariant framework and based on an $SU(N_F)_W$ symmetry
\cite{LM}, associated with a three-system coupling, and used standard
Clebsch-Gordan technology. It was shown to be entirely equivalent to the
covariant approach by Hussain, Korner and Thompson \cite{BS}. In our language
the interactions between an excited meson state ($L$, incoming velocity $v$)
and the two ground states ($L=0$, outgoing velocities $v_1,v_2$) is written
as
$$G_L{\rm Tr}\left[\phi_{\{\mu_1\cdots\mu_L\}}(v)\{\phi(-v_1),\phi(-v_2)\}
     \right] (v_1-v_2)^{\mu_1}\cdots(v_1-v_2)^{\mu_L}, $$
and similarly for baryons; flavour quantum numbers are implied and traced
over as well. In the rest frame of the decaying particle one may easily
recover the expressions of Isgur {\em et al}, since the summation over the
relative momentum indices produces the appropriate $SU(N_F)$ rotation
function \cite{DS}.

Falk and Luke \cite{IWetal} constructed states which are very similar to
ours. However we differ from their approach in that we have recognised
that the heavy quark must be accompanied by other quarks to produce
the correct colourless state; after identifying them and their spin
contributions (which are added to the heavy quark) we have {\em afterwards}
appended the excitation numbers ($L,N$) corresponding to the quark sea and
gluons. The similarity of their spinor wavefunctions with ours can be
established by noticing that the Pauli-Lubanski spin $w_\mu$ acting on a
spinor $\psi(v)$ gives
$$ v^\nu\sigma_{\nu\mu}\gamma_5\psi(v) \rightarrow i(\gamma_\mu + v_\mu)
\gamma_5\psi(v).$$

The connection with Bethe-Salpeter $\Phi$ wavefunctions \cite{BS} is more
distant. In this paper we have adopted the attitude that the excited states
are projected out from the wave-functions using an orthogonal set of
functions of the relative velocities $u$; specifically \cite{WC} for the
mesons say,
\begin{equation}
 \phi(v)_{LM}^N = \int \Phi(v,u){\cal Y}_{LM}^N(u) d^4u,
\end{equation}
with $v^2=1$ or $p^2=m_{NL}^2$ since we are at the meson pole. All we can be
sure of is that the orthogonal functions ${\cal Y}$ contain the spherical
harmonic $Y_{LM}({\bf u})$ in the rest frame $v_0$ of the meson. Naturally,
the dynamics which comes via the Bethe-Salpeter kernel (and is presumably
dominated by non-perturbative gluon exchange) will dictate the remaining
dependence of ${\cal Y}^N$ on $u.v$ and $u^2$, {\em viz.} some linearly
independent combinations of hyperspherical harmonics. An alternative approach
\cite{BS} is the leave the $\Phi$ of initial and final hadrons intact
and carry out an internal integration over projections involving relative
velocities of the current matrix elements.
The only point one can be reasonably sure of in both approaches is that
the degree to which the quarks in the meson are off-shell (or spread in
$|u|$) is of order $\Lambda_{\rm QCD}/m_F$. It could be that a
Bethe-Salpeter equation, something like
\begin{equation}
 [\gamma.(\mu_1 p+k)-m_1]\Phi(p,k)[\gamma.(\mu_2 p-k)+m_2] = \int
 \gamma^\nu\lambda^c.\Phi(p,k').\gamma_\nu\lambda^c D(k-k') d^4k',
\end{equation}
will do the trick, where the exchange propagator $D(k-k')$ is as singular as
$\Lambda^2/(k-k')^4$ at small momentum transfer. Gudehus \cite{BS}
indicates that the details may not be very important anyway in obtaining
the requisite heavy quark symmetry and wavefunctions, and in deriving the
supermultiplet universal form factors, at least for the lowest state $N=0$.
This is a topic that could bear closer investigation.

\acknowledgements

We acknowledge support from the Australian Research Council under
grant number A69231484 and thank Dr. Kreimer for commenting on our work.


\begin{references}

\bibitem[*]{Add} EMail: Bob.Delbourgo@Phys.UTas.Edu.Au and
 Liu@PhysVax.Phys.UTas.Edu.Au
\bibitem{IWetal} N. Isgur and M. Wise, Phys. Rev. Lett. {\bf 66}, 1130
 (1991) and Phys. Rev. {\bf D43}, 819 (1991); N. Isgur, M. B. Wise and
 M. Youssefmir, Phys. Lett. {\bf 254}, 215 (1991); A. Falk and M. Luke,
 Phys. Lett. {\bf 292}, 119 (1992); A. Falk, Nucl. Phys. {\bf B378}, 79
 (1992); S. Balk, F. Hussain, J. G. Korner and G. Thompson, Z. Phys.
 {\bf C59}, 283 (1993); F. Hussain, G. Thompson and J. G. Korner,
 Mainz preprint MZ-TH/93-23 (1993).
\bibitem{BS} T. Gudehus, Phys. Rev. {\bf 184}, 1788 (1969);
 F. Hussain, J. G. Korner and G. Thompson, Ann. Phys. {\bf 206} 334 (1991);
 Y-B Dai, G-S Huang and H-Y Jin, Z. Phys. {\bf C60}, 527 (1993).
\bibitem{DS} R. Delbourgo and A. Salam, Phys. Rev. {\bf 186}, 1516 (1969)
 and Phys. Lett. {\bf 28B}, 497 (1969).
\bibitem{DSS} A. Salam, R. Delbourgo and J. Strathdee, Proc. Roy. Soc. {\bf
 284A}, 146 (1965) and {\em ibid} Proc. Roy. Soc. {\bf 285A}, 312 (1965);
 K. Bardakci, J. M. Cornwall, P. G. O. Freund and B. W. Lee, Phys. Rev.
 Lett. {\bf 14}, 264 (1965);
 M. A. Beg and A. Pais, Phys. Rev. Lett. {\bf 14}, 267 (1965);
 B. Sakita and K. C. Wali, Phys. Rev. {\bf 139}, B1355 (1965).
\bibitem{DRSS} R. Delbourgo, M.A. Rashid, A. Salam and J. Strathdee,
 {\it The \~{U}(12) Symmetry} (IAEA, Vienna, 1965).
\bibitem{MH} E. Eichten and B. Hill, Phys. Lett. {\bf B234}, 511 (1990);
 A. F. Falk, H. Georgi, B. Grinstein and M. B. Wise,
 Nucl. Phys. {\bf B343}, 1 (1990); T. Mannel, W. Roberts and Z. Ryzak,
 Nucl. Phys. {\bf B368}, 204 (1992); J. G. Korner and G. Thompson,
 Phys. Lett. {\bf B264}, 185 (1991).
\bibitem{GR} F. Gursey and L. Radicati, Phys. Rev. Lett. {\bf 299}, 13 (1964);
 B. Sakita, Phys. Rev. {\bf 136}, B1756 (1964); A. Pais, Phys. Rev. Lett.
 {\bf 13}, 175 (1964).
\bibitem{IW} N. Isgur and M. Wise, Phys. Lett. {\bf B232}, 113 (1989);
 {\it ibid} {\bf B237}, 527 (1990).
\bibitem{Georgi} H. Georgi, Phys. Lett. {\bf 240}, 447 (1990) and
 Nucl. Phys. {\bf 348}, 293 (1991)
\bibitem{Melosh} L. L. Foldy and S. A. Wouthuysen, Phys. Rev. {\bf 78},
 29 (1950); S. Tani, Prog. Theor. Phys. {\bf 6}, 267 (1951);
 H. J. Melosh, Phys. Rev. {\bf D9}, 1095 (1974); J. G. Korner and
 G. Thompson, Phys. Lett. {\bf B264}, 185 (1991);
 A. Das, University of Rochester preprint, UR-1325 (1993).
\bibitem{LM} H. J. Lipkin and S. Meshkov, Phys. Rev. Lett. {\bf 14}, 670
 (1965); K. J. Barnes, {\em ibid} {\bf 14}, 798 (1965); D. Horn,
 H. J. Lipkin and S. Meshkov, Phys. Rev. Lett. {\bf 17}, 1200 (1966).
\bibitem{WC} G. C. Wick,  Phys. Rev. {\bf 96}, 1124 (1954);
 R. E. Cutkosky, {\em ibid} 1135 (1954); W. Kummer, Nuovo Cim. {\bf 31},
 219 (1964); S. Mandelstam, Proc. Roy. Soc. {\bf 237A}, 496 (1966);
 R. Delbourgo, A. Salam and J. Strathdee, Phys. Lett. {\bf 21}, 455
 (1966) and Nuovo Cim. {\bf 50}, 193 (1967).
\end{references}
\end{document}